\documentclass[12pt]{iopart}

\usepackage{iopams}
\usepackage{color}
\usepackage{subfigure}
\usepackage{epsfig}

\newcommand{\la}{\lambda}

\newcommand{\ga}{\gamma}

\newcommand{\be}{\begin{equation}}
\newcommand{\ee}{\end{equation}}
\newcommand{\bea}{\begin{eqnarray}}
\newcommand{\eea}{\end{eqnarray}}
\newcommand{\nn}{\nonumber}

\begin{document}
\title[Superstatistical Wishart-Laguerre Ensembles]{Superstatistical
generalisations of Wishart-Laguerre ensembles of random matrices}

\author{A.Y. Abul-Magd$^1$, G. Akemann$^2$, and P. Vivo$^3$}
\address{\it $^1$Faculty of Engineering Sciences, Sinai University, El-Arish,
  Egypt\\
$^2$Department of Mathematical Sciences
\& BURSt Research Centre,\\
\ Brunel University West London,
UB8 3PH Uxbridge, United Kingdom\\
$^3$Abdus Salam International Centre for Theoretical Physics\\
\ Strada Costiera 11, 34014 Trieste, Italy
}



\begin{abstract}
Using Beck and Cohen's superstatistics, we introduce in a
systematic way a family of generalised Wishart-Laguerre ensembles
of random matrices with Dyson index $\beta$ = 1,2, and 4. The
entries of the data matrix are Gaussian random variables whose
variances $\eta$ fluctuate from one sample to another according to
a certain probability density $f(\eta)$ and a single deformation
parameter $\ga$. Three superstatistical classes for $f(\eta)$ are
usually considered: $\chi^2$-, inverse $\chi^2$- and
log-normal-distributions. While the first class, already considered
by two of the authors, leads to a power-law decay 
of the spectral density, 
we here introduce
and solve exactly a superposition of Wishart-Laguerre
ensembles with inverse $\chi^2$-distribution. The corresponding
macroscopic spectral density is given by a $\ga$-deformation of the
semi-circle and Mar\v{c}enko-Pastur laws, on a non-compact support
with exponential tails. After discussing in detail the validity of Wigner's surmise in the Wishart-Laguerre class,
we introduce a generalised $\gamma$-dependent surmise with stretched-exponential tails, which
well approximates the individual level spacing distribution in the bulk. 
The analytical results are in excellent agreement with
numerical simulations. 
To illustrate our findings we compare the $\chi^2$- and inverse $\chi^2$-class
to empirical data from financial covariance matrices.

\end{abstract}
\date{\today}


\section{Introduction}

Random matrix theory (RMT) is known to find applications in many
physical systems \cite{guhr,mehta}. Its central assumption is that
the Hamiltonian of the system under consideration can be replaced
by an ensemble of random matrices that are consistent with its
global symmetries. The matrix elements of such a random
Hamiltonian are typically independent Gaussian variables with mean
zero and variance one in the real, complex or quaternion domain. These are respectively
labelled by the Dyson index $\beta$ = 1, 2 and 4, which in turn corresponds to 
the invariance group of the ensemble (orthogonal, unitary or symplectic). 
Another set of ensembles
of random matrices, which is well studied in RMT is known as
Wishart-Laguerre or chiral ensemble and will be denoted by WL in
the following. The WL ensemble contains random matrices of the form \cite{wishart} $\mathbf{W=X^{\dagger}X}$, where
$\mathbf{X}$ is a rectangular matrix of size $M\times N~$($M>N$),
whose entries are independent Gaussian variables in the simplest
case, and $\mathbf{X^{\dagger}}$ is the Hermitian conjugate of
$\mathbf{X}$. As such, the WL ensemble contains (positive
definite) covariance matrices $\mathbf{W}$ of maximally random
data sets. They have since appeared in many different contexts
ranging from mathematical statistics, statistical physics and
quantitative finance to gauge theories, quantum gravity and
telecommunications \cite{app}.

The Gaussian distribution of matrix elements above can be obtained
by maximising the Boltzmann-Gibbs-Shannon entropy subject to the
constraint of normalisation and of fixed expectation value of
Tr$\left( \mathbf{X}^{\dagger}\mathbf{X}\right)$ \cite{balian}.
This observation has been taken by several authors \cite{toscano}
as a starting point for generalising the classical Wigner-Dyson (WD) ensembles,
by maximising Tsallis' non-extensive entropy
subject to the same constraints. The resulting distributions of
matrix elements are no longer Gaussian but rather follow a power law
(for a different generating mechanism of such ensembles see \cite{bohigas}).
Recently, a similar generalisation of the WL ensembles has been
worked out \cite{AV}, following earlier studies \cite{burda} on
the covariance matrices of financial data. 

In this context, it is
known \cite{biroli} that the spectral density of covariance
matrices built from empirical data may deviate significantly from
their purely random (WL) counterpart: in order to obtain a better
agreement, it is necessary to introduce correlations among the
matrix elements of $\mathbf{X}$, typically allowing the variance
of each entry to fluctuate. 
In \cite{biroli} this is done using a multivariate Student distribution, where
the random entries of $\mathbf{X}$ are written as a product of two variables with a Gaussian 
and {\it inverse $\chi^2$}-distribution respectively.
Generically, this approach spoils the
complete solvability of the model and prevents from going beyond
the average spectral density \cite{biroli}. In order to study
other correlations, it is therefore of great interest to introduce
generalisations of WL where the independence of
$\mathbf{X}$-entries is dropped, but the exact solvability is
retained. One of the first examples of such models was provided in
\cite{AV}, where a good fit to the power-law decay of financial
covariance spectra was obtained.

It is desirable to place the model \cite{AV} within a more general
framework, as was done previously for the WD class
\cite{sust1,muttalib}. The key observation is to resort to the
ideas
of superstatistics (or statistics of a statistics) proposed by
Beck and Cohen \cite{BC}. Outside RMT, this formalism has been
elaborated and applied successfully to a wide variety of physical
problems, e.g., in \cite{cohen}. In thermostatics, superstatistics
arises as a weighted average of ordinary Boltzmann statistics due
to fluctuations of one or more intensive parameters (e.g. the
inverse temperature). Typically, the distribution of the
superstatistical parameter falls into three 'universality'
classes: $\chi^2$-, inverse $\chi^2$- and log-normal distributions
(see section \ref{three} for a more detailed discussion).
Superstatistical RMT \cite{sust1} analogously assumes that
the Hamiltonian of the system is locally described by a standard WD
ensemble with a given variance, and when averaging over the whole
system the variance is integrated over with a specific
distribution. Consequently, superstatistical RMT is a
superposition or integral transform of the usual Gaussian RMTs.

The same concept of integral transforms of standard RMT has also
appeared in other contexts, e.g. the fixed or restricted trace
ensembles \cite{ACMV}, also called norm-dependent ensembles. Very
general results can be deduced for them, without specifying the
distribution \cite{muttalib,Guhr}. However, no systematic
applications of superstatistics and integral transforms to the WL
class are known so far. It is the purpose of this paper to fill this gap. 
We are going to
introduce first a family of generalised WL ensemble depending on a single real parameter $\ga$, where the
prescriptions of superstatistics are incorporated in a systematic
way, and then solve exactly a representative example where the
entries of $\mathbf{X}$ are Gaussian random variables whose variances
fluctuate from sample-to-sample according to an inverse $\chi^2$-distribution. 
Given the appearance of such a distribution in modelling the volatility of financial markets \cite{biroli},
it is sensible to compute the tail of the spectral density in our model and compare it
to the findings in \cite{biroli} and 
\cite{AV}. However, because of WL ensembles being tailored to times series of data, we expect
applications beyond financial covariance matrices as applied here.

Despite
the complicated correlations among the entries, and thanks to the integral transform
mapping we can go to an eigenvalue basis and write the joint
probability density function (jpdf) of eigenvalues as an integral
over the corresponding WL one, a feature that was already
exploited in \cite{AV} for the $\chi^2$-distribution. 
Being able to perform all integrals analytically, we have full control over
exchanging the large $N$-limit with our deformation parameter $\ga$. In the limit $\ga\to\infty$,
our generalised ensembles are designed to recover the standard WL, and this fact will constitute an important
consistency check in the following.
A third
class of models can be obtained by folding with the log-normal
distribution, but we will not deal with this case here, lacking a full
analytic solution.

It is worth mentioning that the applicability of this formalism is
not limited to the above three universality classes. In particular, it
would be very interesting to investigate whether the choice of the
best superstatistical distribution could be inferred from the data
set, instead of being somehow 'postulated' a priori. This would be
reminiscent of the Bayesian approach to superstatistics
\cite{sattin}, this time in the more complicated RMT setting.

In the next section \ref{superWL}, we briefly review the
prescriptions of superstatistical RMT, together with the three
universality classes in subsection \ref{three}. Then we give a
step-by-step derivation of our superstatistical WL ensemble in
subsection \ref{building}. In section \ref{setup} we define the
spectral properties to be computed in our model and discuss their
universality. We then
investigate the large matrix size limit of the spectral density in
section \ref{density}, where we distinguish between square and
rectangular matrices of size $N=cM$ in the two subsections. 
In section \ref{spacing} we first present a detailed discussion on the applicability of the standard Wigner's
surmise for the level spacing in the WL class, and then we derive a new, generalised $\gamma$-dependent
surmise which is appropriate to describe with excellent approximation the individual level spacing in the bulk
of our superstatistical model.
Numerical checks on the analytical results are provided throughout the text and the algorithm used is described in
\ref{numerical}. 
In section \ref{apply} we compare two superstatistical distributions to empirical data from 
financial covariance matrices, 
before offering concluding remarks in
section \ref{conclusio}.


\section{Superstatistical Wishart-Laguerre ensembles}\label{superWL}

The probability density of the data-matrix entries in WL ensembles
is given by
\begin{equation}
\label{PWL} P_{{W\!L}}\left(  \eta;\mathbf{X}\right)
=\frac{1}{Z_{W\!L}(\eta)}\exp\left[ -\eta \beta\Tr(
\mathbf{X}^{\dagger}\mathbf{X})  \right]  ,
\end{equation}
where $\eta$ is proportional to the inverse variance of matrix
elements.

The normalisation is given by the partition function \be
\label{ZWL} Z_{W\!L}(\eta)=\int d\mathbf{X}\exp\left[ -\eta \Tr(
\mathbf{X}^{\dagger }\mathbf{X}) \right] =\frac{1}{\beta NM}\left(
\frac{\pi}{\eta \beta}\right)^{\beta MN//2} \ . \ee As mentioned
above, \mbox{\bf X} is a matrix of size $M\times N~$ with real,
complex or quaternion real elements for the values $\beta$ = 1, 2
or 4, respectively. We parametrise $N=cM$ for later convenience,
where $c\leq1$ distinguishes two different large-$N$ limits. The
integration measure $d\mathbf{X}$ is defined by integrating over
all independent matrix elements of \mbox{\bf X} with a flat
measure. The statistical information about the positive definite
eigenvalues of the Wishart matrix $\mathbf{W=X^{\dagger}X}$, or
equivalently the singular values of the matrix $\mathbf{X}$ can be
obtained integrating out all the undesired variables from the
distribution of the matrix elements, using its orthogonal, unitary
or symplectic invariance for $\beta=1,2$ and 4 respectively.

We now define a  superstatistical family of WL ensembles. The key
ingredients are the following:
\begin{enumerate}
  \item A real deformation parameter $\ga>0$, which roughly quantifies how
far the new model lies from the traditional, unperturbed WL
ensemble. In the limit $\ga\to\infty$, we expect to recover WL
exactly.
  \item A normalised probability density $f(\eta)$, such that
  \be
1=\int_0^\infty d\eta\, {f}(\eta)\ . \ee It is understood that
$f(\eta)$ depends on $\ga$ as well, but we will not show this
dependence explicit in order to keep the notation light.
\end{enumerate}

The probability distribution of matrix elements for the
generalised model is then obtained as follows:
\begin{equation}
P(\mathbf{X})=\int_{0}^{\infty}d\eta\ f(\eta)P_{W\!L}\left(
\ell(\ga)\eta;\mathbf{X}\right)=\Big\langle P_{W\!L}\left(
\ell(\ga)\eta;\mathbf{X}\right)\Big\rangle_{f}  . \label{PCB}
\end{equation}
where $\langle\cdot\rangle_f$ means average over the distribution
$f(\eta)$ and $\ell(\ga)$ is a simple function of the deformation
parameter (see subsection \ref{building} below for details). The
choice of distribution $f(\eta)$ is determined by the system under
consideration. In the absence of fluctuations of the variance,
$f(\eta)=\delta(\eta-\eta_{0})$ and we reobtain the standard WL
ensemble. Typically, the distribution $f(\eta)$ depends
explicitly on the parameter $\ga$, in such a way that for
$\ga\to\infty$ a delta-function limit is obtained and thus the WL
results (before or after taking $N$ large) are duly recovered.

While the distribution defined in eq. (\ref{PCB}) is formally
normalised, when exchanging the integration $\int d\mathbf{X}$
with $\int d\eta$, the prescription given so far is not complete.
The choice of a distribution $f(\eta)$, and in particular of the
$N$-dependence of its parameter(s) has to be such that a) the
integral over $P(\mathbf{X})$ is convergent, and that b) an $N$
independent limit for the spectral density can be found after a
proper rescaling of variables. We will come back to this issue
below.

\subsection{The three superstatistical classes}\label{three}

Beck
et al. \cite{bcs} have argued that experimental data can be described by
one of three superstatistical universal classes, namely the $\chi^{2}$-,
inverse $\chi^{2}$-, or log-normal-distribution. Below we briefly discuss each
class and the properties of its
corresponding distribution, before turning to the detailed
solution for the inverse $\chi^{2}$-class in the next sections.

\begin{itemize}

\item[1)] \underline{\bf $\chi^{2}$-distribution:}

The
$\chi^{2}$-distribution with degree $\nu$ and average $\eta_0$ is given by
\begin{equation}\label{chi22}
f_1(\eta)=\frac{1}{\Gamma\left(\frac\nu2\right)}
\left(  \frac{\nu}{2\eta_{0}}\right)  ^{\frac12\nu}
\eta^{\frac12\nu-1}\exp\Big[-\nu\eta\frac{1}{2\eta_{0}}\Big],
\end{equation}
where we define
$\eta_{0}=\int_{0}^{\infty}\eta f(\eta)d\eta\equiv \langle\eta\rangle$.
This distribution is appropriate if the variable $\eta$ can be represented
as a sum of squares of $\nu$ Gaussian random variables.

The superstatistical distribution arising from (\ref{chi22}) is
Tsallis' statistics with power law tails \cite{tsallis} and is
believed to be relevant e.g. for cosmic ray statistics \cite{beckray}.

\item[2)] \underline{\bf inverse $\chi^{2}$-distribution:}

This distribution is found if $\eta^{-1},$
rather than $\eta,$ is the sum of several squared
Gaussian random variables.
Its distribution $f_2(\eta)$ is given by
\begin{equation}
\label{inversechi}
f_2(\eta)=\frac{\eta_{0}}{\Gamma\left(\frac\nu2\right)}
\left(  \frac{\nu\eta_{0}}{2}\right)^{\frac12\nu}
\frac{1}{\eta^{\frac12\nu+2}}\exp\Big[-\nu\eta_{0}\frac{1}{2\eta}\Big],
\end{equation}
with degree $\nu$ and average $\eta_0$.

This class is appropriate for systems exhibiting velocity
distribution with exponential tails \cite{touchette} and was shown
to describe the spectral fluctuations of billiards with mixed
regular-chaotic dynamics better than the other two distributions
\cite{sust2}.

\item[3)] \underline{\bf $\log-$normal distribution:}

In the third class, instead of being the sum of contributions, the
random variable $\eta$ may be generated by a multiplicative random
processes. Then its logarithm $\log(\eta)=\sum_{i=1}^{\nu}\log
(x_{i})$ is a sum of $\nu$ Gaussian random variables $x_{i}$. Thus
it is log-normally distributed:
\begin{equation}
f_3(\eta)=\frac{1}{\sqrt{2\pi}\ v\eta}\exp\left[-\left[
    \log(\eta/\mu)\right]^{2}
\frac{1}{  2v^{2}}\right].
\end{equation}
It has an average $\langle\eta\rangle=\mu\sqrt{w}$
and variance $\mu^{2}w(w-1)$, where
$w=\exp(v^{2})$.

This class has been found relevant for Lagrangian and Eulerian
turbulence \cite{cohen,mordant}.

Because apparently this class does not lead to closed analytical
results for the distribution of matrix elements $P_3(\mathbf{X})$
or its correlation functions we do not use it as an example in
this paper.

\end{itemize}

\subsection{Building the superstatistical ensembles}\label{building}
Since the prescription (\ref{PCB}) is not completely
straightforward, we provide here a step-by-step construction of
the superstatistical probability density
$P\left(\mathbf{X}\right)$. 
We focus here on 
the inverse $\chi^2$-distribution (\ref{inversechi}) as a new example.
\begin{enumerate}
  \item We start with the usual WL probability density for the
  entries, eq. (\ref{PWL}):
\begin{equation}
\label{PWL2} P\left(\mathbf{X}\right) \rightsquigarrow\exp\left[
-\eta\beta\Tr( \mathbf{X}^{\dagger}\mathbf{X})  \right]\ .
\end{equation}
Henceforth we omit the normalisation constants.
  \item Next, we convolve the WL weight with the inverse $\chi^2$-distribution
(\ref{inversechi}):
\begin{equation}
\label{PWL3} P\left(\mathbf{X}\right)
\rightsquigarrow\frac{1}{\eta^{\frac12\nu+2}}
\exp\Big[-\nu\eta_{0}\frac{1}{2\eta}\Big]
\exp\left[-\eta\beta\Tr( \mathbf{X}^{\dagger}\mathbf{X})  \right]\ .
\end{equation}
  \item The $\nu$ parameter
becomes the deformation parameter of the model as $\gamma=\nu/2$.
   For a clearer notation, we now make the further replacement
   $\eta=\nu\eta_0\xi/2$ and eventually integrate over the
   possible values for $\xi$:
\begin{equation}
\label{PWL4} P\left(\mathbf{X}\right)
\rightsquigarrow\int_0^\infty d\xi
\frac{1}{\xi^{\ga+2}}\exp\left(-\frac{1}{\xi}\right)\exp\left[
-\xi\ga\beta\Tr( \mathbf{X}^{\dagger}\mathbf{X})  \right]\ .
\end{equation}
Note that the average $\eta_0$ is no longer needed explicitly and
has been absorbed in the other parameters. In (\ref{PWL4}), the
deformation parameter $\ga$ appears in both the superstatistical
weight function and inside the WL Gaussian weight (here
$\ell(\ga)=\ga$, while in the \cite{AV} model $\ell(\ga)=1/\ga$).
\item Although formally exact, eq. (\ref{PWL4}) does not
lead to a $N$-independent spectral density,
as can be quickly realized with a modest amount of foresight (see
(\ref{zga}) and (\ref{Zratio})). The last step is thus to amend
slightly (\ref{PWL4}), including a suitable extra factor for normalisation
$\xi^{-(\beta/2)NM}$:
\begin{equation}
\label{PWL5}
\fl P_2\left(\mathbf{X}\right) \equiv\int_0^\infty d\xi
\frac{1}{\xi^{\ga+2-(\beta/2)MN}}\exp\left(-\frac{1}{\xi}\right)\exp\left[
-\xi\ga\beta\Tr( \mathbf{X}^{\dagger}\mathbf{X})  \right]\ .
\end{equation}
\item Eventually, the integral in (\ref{PWL5}) can be evaluated,
and gives the probability density for the entries of our
superstatistical WL model as:
\begin{equation}\label{PWL6}
\fl P_2\left(\mathbf{X}\right)\propto \Big(
\Tr(\mathbf{X}^{\dagger}\mathbf{X})\Big)^{\frac12(\ga+1-\beta
NM/2)} K_{\ga+1-\beta NM/2}
\left(2\sqrt{\beta\ga\Tr(\mathbf{X}^{\dagger}\mathbf{X})}\right).
\end{equation}
Here $K_\ga(z)$ is the modified $K$-Bessel function of second
kind.
\end{enumerate}
This distribution is the main subject of this paper. It is
convergent for all values of $\ga>0$, and we will show later that
our choice of the normalisation $\xi^{-(\beta/2)NM}$ effectively
leads to an $N$-independent limit for the average spectral density
at fixed and finite $\ga$.

We can analytically recover the standard WL ensemble for
$\ga\to\infty$ (at fixed and finite $N,M$) using the following
{\it non-standard asymptotic} of the Bessel-$K$ function\footnote{Note
that the argument
  $\sqrt{z}$ of the
  Bessel-$K$ function becomes quadratic in the exponent.}:
  \be
z^{\frac\ga2}K_\ga(2\sqrt{\ga z})\sim \sqrt{\frac{\pi}{2}}
\ga^{\frac12(\ga-1)}\e^{-\ga}\exp[-z] \ . \label{Kasymp} 
\ee 
While we did not find eq. (\ref{Kasymp}) in tables,
it follows easily form the following integral representation 8.432.6
\cite{Grad}, after 
using a saddle
point approximation
\be K_\ga(x)\ =\
\frac{x^\ga}{2^{\ga+1}}\int_0^\infty dt\
t^{-\ga-1}\exp\left[-t-\frac{x^2}{4t}\right]\ ,\ \ x^2>0\ .
\label{Krep} \ee
To derive eq. (\ref{Kasymp})
we have included the fluctuations around the saddle point and
used the standard notation $f\sim g$ to mean that $f/g\to 1$. The
$\ga$-dependent prefactor can be cancelled by a proper
normalisation. 

It is important to stress that the choice of the normalisation
$\xi^{-(\beta/2)MN}$ does not affect the correct $\ga\to\infty$
asymptotics in any way, while being the only sensible prescription
when taking the large $N,M$ limit at fixed $\ga$ for the average
spectral density (see section \ref{density}).

It is also worth mentioning that, following the same steps as
above, one can incorporate the other two superstatistical
distributions. For example, for the $\chi^2$-distribution, one is
naturally led to: \bea P_1(\mathbf{X})&\propto& \int_0^\infty
d\xi\ \xi^{\ga-1+\frac12\beta NM}
\e^{-\xi}\exp\Big[-\frac{\xi\beta}{\ga}
\Tr(\mathbf{X}^{\dagger}\mathbf{X})\Big]
\nn\\
&\propto & \left( 1+\frac{\beta}{\ga}
\Tr(\mathbf{X}^{\dagger}\mathbf{X})\right)^{-\ga+\frac12\beta
  NM}.
\label{P1} \eea The power-law decay of the matrix elements
translates into all correlation functions \cite{AV}, depending on
a single rescaled parameter, $\alpha=\ga-\frac\beta2 NM-1>0$, which is kept
fixed in the large-$N$ limit. The
reduction back to standard WL can be made both before (and after)
the large-$N$ limit using
\be
\lim_{\ga\to\infty}(1+\ga^{-1}z)^{-\ga}\ =\ \e^{-z}\ .
\ee
For a more detailed discussion we refer to \cite{AV}.

\section{Generalisation of WL with exponential tails}\label{setup}

After introducing the probability distribution of matrix elements
in our generalised inverse $\chi^2$-WL eq. (\ref{PWL5}) we go to
an eigenvalue basis in this section and define all correlation
functions.

The corresponding joint probability distributions function (jpdf)
of positive definite eigenvalues $\lambda_{1},...,\lambda_{N}$ of
the matrix $\mathbf{W}=\mathbf{X}^\dag \mathbf{X}$ reads \bea
\label{PiXdef} &\fl{\cal P}_{\ga}
(\lambda_{1},...,\lambda_{N})\propto \int_{0}^{\infty}d\xi\,
\frac{1}{\xi^{\ga+2-\frac\beta2NM}} \exp\Big[ -\frac1\xi\Big]
{\cal P}_{W\!L}(\lambda_{1},...,\lambda_{N};\xi)\\
&\fl\propto \prod_{i<j}^N|\la_j-\la_i|^\beta
\prod_{k=1}^N\la_k^{\frac\beta2(M-N+1)-1} \left(
\sum_{i=1}^N\la_i\right)^{\frac{1}{2}(\ga+1-\beta N M/2)}
K_{\ga+1-\beta NM/2}
\left(2\sqrt{\beta\ga\sum_{j=1}^N\la_j}\right). \nn \eea Here we
have used the jpdf  ${\cal P}_{W\!L}$ of the standard WL ensemble,
which is given by
\begin{equation}
\label{PWLev} {\cal
P}_{W\!L}(\lambda_{1},...,\lambda_{N};\xi)\propto
{\displaystyle\prod\limits_{i<j}^{N}} \left\vert
\lambda_{j}-\lambda_{i}\right\vert ^{\beta}
{\displaystyle\prod\limits_{k=1}^{N}}
\lambda_{k}^{\frac{\beta}{2}(M-N+1)-1}
\e^{-\xi\ga\beta\sum_{l=1}^{N}\lambda_{l}}.
\end{equation}
For convergence $\ga$ has to be positive. Both jpdf's still have
to be normalised by the respective partition function $Z_\ga$ and
$Z_{W\!L}(\xi)$, obtained by integrating over all eigenvalues.

The $k$-point density correlation functions defined by integrating
out $N-k$ arguments of the jpdf are given as:
\begin{equation}\label{Corrfun}
R_{\ga,\,k}(\lambda_{1},...,\lambda_{k})\equiv\frac{N!}{\left(  N-k\right)  !}
\frac{1}{Z_\ga}\int_{0}^{\infty}d\lambda_{k+1}\cdots\int_{0}^{\infty}
d\lambda_{N}{\cal P}_{\ga}(\lambda_{1},...,\lambda_{N}).
\end{equation}
Because of the linear relationship between the jpdf of our
ensemble and WL, eq. (\ref{PiXdef}), we can express the $k$-point
functions of the former through the latter: \be
R_{\ga,k}(\lambda_{1},...,\lambda_{k})=\int_{0}^{\infty} d\xi\,
\frac{1}{\xi^{\ga+2-\frac\beta2NM}}\ \e^{-\frac1\xi}\
\frac{Z_{W\!L}(\xi)}{Z_\ga}
R_{W\!L,\,k}(\lambda_{1},...,\lambda_{k};\xi), \label{RRrel} \ee
where $R_{W\!L,\,k}(\lambda_{1},...,\lambda_{k};\xi)$ is the
corresponding $k$-point function for the WL ensemble defined in
the same fashion. The $\xi$-dependent ratio of the two partition
functions which are also linearly related through 
\be Z_\ga\ =\
\int_{0}^{\infty} d\xi\, \frac{1}{\xi^{\ga+2-\frac\beta2NM}}\
\e^{-\frac1\xi}\ Z_{W\!L}(\xi)\ , 
\label{zga}
\ee 
easily follows
from eq. (\ref{ZWL}): 
\be \frac{Z_{W\!L}(\xi)}{Z_\ga}\ =\
\frac{\xi^{-\frac\beta2 NM}}{ \Gamma(\ga+1)}\ . 
\label{Zratio} 
\ee
Comparing eq. (\ref{Zratio}) and (\ref{RRrel}), it is already
apparent that the choice of the normalisation in eq. (\ref{PWL4})
was in fact necessary to neutralise exactly the $N$- and $M$-dependence
coming from the partition functions.
In all our formulas, the limit $\ga\to\infty$ at finite $N$ and
$M$ leads back to WL, using eq. (\ref{Kasymp}).

\subsection{Universality}\label{univ}

We close this section by discussing the universality of generalisations of WL. 
Following from a variational, generalised entropy
principle we should expect a certain robustness of our results, which 
is indeed the case. 

First, all $k$-point
correlation functions of the WL ensembles are explicitly known for
finite and infinite-$N$, in terms of Laguerre polynomials and
their asymptotics. Thanks to the integral mapping eq.
(\ref{RRrel}), the superstatistical models are exactly solvable as
well (see e.g. \cite{muttalib,AV}). Also because of this, we could
allow for a more general WL ensemble with a polynomial
potential $V$ instead of the Gaussian,
$\Tr(\mathbf{X}^{\dagger}\mathbf{X})\to \Tr
V(\mathbf{X}^{\dagger}\mathbf{X})$, in order to probe the universality of our
results under deformations. 
These non-Gaussian ensembles
can again be solved using the technique of (skew-) orthogonal
polynomials for any finite $N$ and $M$, and again the complete
solvability of the superstatistical generalisation is guaranteed
by the very same integral relation eq. (\ref{RRrel}).

It is known that in the microscopic large-$N$ limit the asymptotic of the 
orthogonal polynomials for these non-Gaussian models is universal 
\cite{ADMN,VS}
(for rigorous mathematical 
proofs see \cite{universal} and references therein).
This implies that the microscopic correlations in our model
remain unchanged after the integral transform. As an example for such a
microscopic quantity we deal with the level spacing distribution in section \ref{spacing}. 
For a detailed
discussion of the universal microscopic spectral correlations 
with a $\chi^2$-distribution
we refer to \cite{AV}, including references. 

About the macroscopic large-$N$ limit, the (generalised) semi-circle and 
Mar\v{c}enko-Pastur densities are known to be less robust. They are altered by  
deformations via a polynomial potential $V$ but still remain
computable, see e.g. the discussion in \cite{ACMV} 
and references therein.  However, a different kind of deformation is known that 
leaves the semi-circle or 
Mar\v{c}enko-Pastur density unchanged: the so-called Wigner ensembles (see 
\cite{Bai} for a review). The Gaussian random variables of the WL
ensemble are replaced by independent random variables with zero mean and fixed
second moment. While this generalisation destroys both the invariance as well
as the integrability of higher correlation functions, its large-$N$ macroscopic
density is the same as in WL, and thus also our integral transform of WL, after taking the large-$N$ limit. 



\section{The macroscopic spectral density}\label{density}

In this section we focus on the simplest observable, the spectral
density $R_{\ga,k=1}(\la)$ obtained by integrating out all
eigenvalues but one\footnote{Following the general definition eq.
(\ref{Corrfun}), $R(\la)$ denotes the spectral density
normalised to $N$.}.

When taking the large-$N,M$ limit, we keep the combination
$c=N/M\leq 1$ fixed. This leads to two different behaviours for
the WL spectral density: the semi-circle (in squared variables) for $c=1$ and the
Mar\v{c}enko-Pastur (MP) density for $c<1$, see eqs.
(\ref{MPc=1}), (\ref{MPc<1}) below. Therefore we expect two
different limits for our generalised ensembles as well, and the
generalisations of these two well-known WL results will be dealt
with in two separate subsections below.

Although the average spectral density for WL (and hence for our
model as well through eq. (\ref{RRrel})) is known explicitly for
any finite-$N,M$ in terms of Laguerre polynomials (see e.g.
\cite{mehta,AV}), the generalised spectral density for large $N$
can be obtained following a much simpler route: we replace the
finite-$N$ WL quantity under the integral by its large-$N$ result.
The correctness of this approach has already been shown in
\cite{AV}. As a further check, we can eventually take
$\ga\to\infty$ to recover the semi-circle or MP density.


\subsection{Generalised semi-circle for law $c$ = 1}\label{semicirc}

\begin{center}
\begin{figure*}[htb]
  \unitlength1.0cm
  \epsfig{file=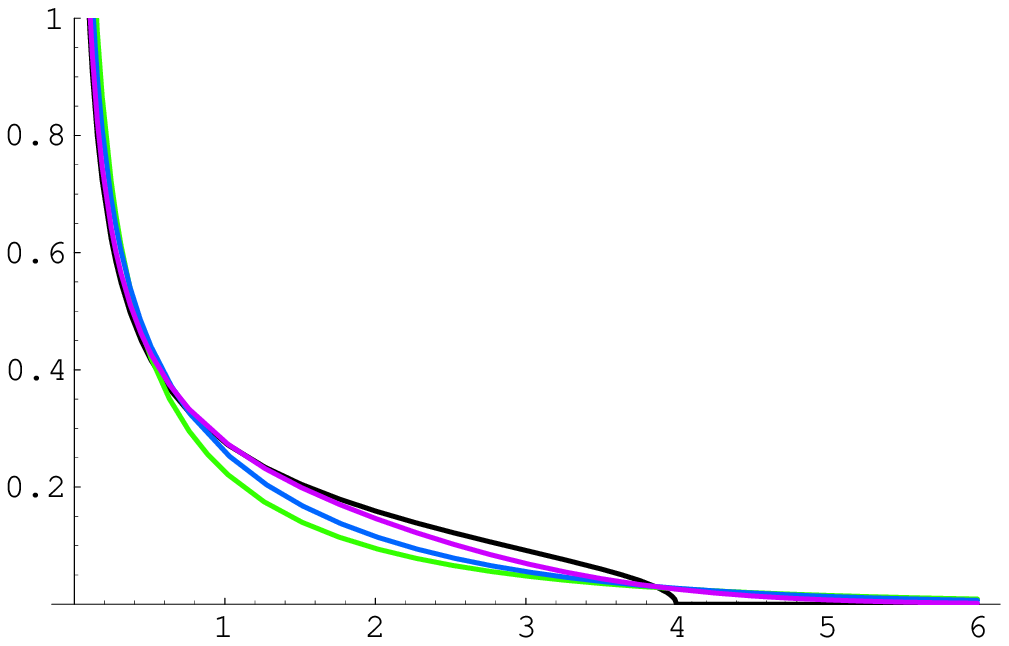,clip=,width=7.6cm}
  \epsfig{file=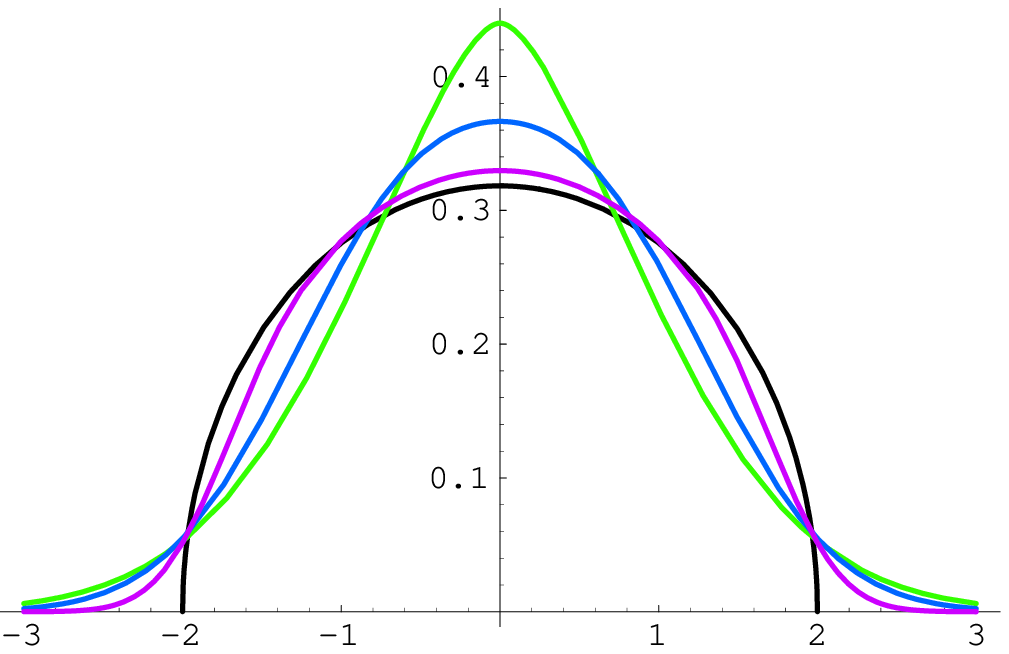,clip=,width=7.8cm}
  \caption{
    \label{DensC1}
The generalised semi-circle eq. (\ref{rhogaU}) before (left)
and after (right)
the mapping eq. (\ref{Rmap}) to $\mathbb{R}$ for different values of
$\ga=0.5,2,10$ (green, blue, violet respectively). For comparison,
the standard WL semicircle is plotted as well (black). }
\end{figure*}
\end{center}

In the case $c=1$, the large-$N$ asymptotic expression for the
density of eigenvalues of a WL ensembles with distribution
$\exp[-\eta\beta\Tr\mathbf{X}^\dagger \mathbf{X}]$ eq. (\ref{PWL})
is given by \be \label{densityWishart}
\lim_{N\gg1}R_{W\!L,\,1}(\la)=
\frac{\eta}{\pi}\sqrt{\frac{2N}{\eta\la}-1} \ \ \mbox{with}\ \
\la\in(0,2N/\eta]\ , \ee and 0 otherwise. An $N$-independent
macroscopic density with average one is obtained by rescaling
$\la\to x\ \langle\la\rangle_{W\!L}$ with its mean
$\langle\la\rangle_{W\!L} =\langle\frac1N\Tr\mathbf{X}^\dagger
\mathbf{X}\rangle_{W\!L}=\frac{M}{2\eta}$, calculated with respect
to the above distribution. We obtain
\begin{equation}
\label{MPc=1}
\rho_{W\!L}(x)\ \equiv\   \lim_{N\to\infty}\frac1N\langle \lambda\rangle
R_{W\!L,\,1}(x\langle \lambda\rangle)\ =\ \frac{1}{2\pi}\sqrt{
\frac{4}{x}-1}\ ,\ \ \mbox{with}\ \ x\in(0,4]\ .
\end{equation}
It diverges as $1/\sqrt{x}$ at the origin and vanishes as a square
root at the upper edge of support. Eq. (\ref{MPc=1})
corresponds to a semi-circle after mapping eigenvalues from
$\mathbb{R}_+$ to $\mathbb{R}$, and we will comment more on that
below (see Fig. \ref{DensC1}).

Following the same procedure as above for the inverse $\chi^2$ ensemble,
we first need to determine the average eigenvalue in our ensemble (see e.g.
\cite{AV} appendix A):
\be
\langle\la\rangle_\ga\ \equiv\ \frac{1}{Z_\ga}
\int_{0}^{\infty} d\xi\, \frac{1}{\xi^{\ga+2-\frac\beta2NM}}\
\e^{-\frac1\xi}\ Z_{W\!L}(\xi)\langle \lambda(\xi)\rangle
\ =\ \frac{M(\ga+1)}{2\ga}\ ,
\label{lagavev}
\ee
using eq. (\ref{Zratio}) and the mean in WL above.

Next we define the generalised macroscopic $N$-independent density as
\bea
\rho_\ga(x) &\equiv& \lim_{N\to\infty}\frac1N\langle
\lambda\rangle_\ga R_\ga(x\langle \lambda\rangle_\ga)
\nn\\
&=&  \lim_{N\to\infty} \frac1N\langle \lambda\rangle_\ga
\int_\mathcal{I} d\xi\ \frac{1}{\xi^{\ga+2-\frac\beta2 NM} }\
\e^{-\frac1\xi} \frac{Z(\xi)}{Z_\ga} \frac{\xi\ga}{\pi}\
\sqrt{\frac{2N}{\xi\ga x\langle \lambda\rangle_\ga}-1}\ .
\label{rhoga1} \eea Here we have simply inserted the large-$N$ WL
density eq. (\ref{densityWishart}) with the parameter
$\eta=\xi\ga$ into the integrand eq. (\ref{RRrel}). Consequently
the interval of integration is truncated and given by
$\mathcal{I}=(0,\frac{4}{x(\ga +1)}]$. Thanks to the correct choice of
normalisation in eq. (\ref{PWL5}), the $N$-dependence completely
drops out and for any fixed $\ga>0$ the new spectral density admits the
following integral representation, after changing variables
$t=\frac{4}{\xi x(\ga+1)}-1$:
\be
\fl \rho_\ga(x)=\frac{(\ga+1)^{\ga+1}}{2\pi\Gamma(\ga+1)}\
\left(\frac{x}{4}\right)^{\ga}\int_0^\infty dt\
\exp\left[-\frac{(\ga+1)}{4}\,x(t+1)\right]\, (t+1)^{\ga-1}\ \sqrt{t}\ .
\label{rhoga} \ee
This is our first main result of this section.
This density as well as its first moment are correctly normalised
to $1$:
\begin{equation}\label{Normalization}
  \int_0^\infty \rho_\ga(x)dx=1=\int_0^\infty x\rho_\ga(x)dx\ .
\end{equation}

As a consistency check, we can now take the limit $\ga\to\infty$.
The integral becomes amenable to a saddle point approximation, and
taking into account the fluctuations around the saddle point
we reproduce exactly eq. (\ref{MPc=1}) as we should. Eq.
(\ref{rhoga}) is plotted in Fig. \ref{DensC1} (left) for various values
of $\ga$. Our generalised density has support on the full
$\mathbb{R}_+$, in contrast with the compact support of the WL
semi-circle.

In order to  analytically derive the asymptotic behaviour of our
generalised semi-circle eq. (\ref{rhoga}) at the origin $x\to0$
and at infinity $x\to\infty$ it is more convenient to express the
integral through special functions. Using eq. 3.383.5 \cite{Grad}
we can express the density eq. (\ref{rhoga}) as 
\be
\label{rhogaU}
\fl \rho_\ga(x)=\frac{(\ga+1)^{\ga+1}}{4\sqrt{\pi}\Gamma(\ga+1)}\
\exp\left[-(\ga+1)\frac{x}{4}\right]\left(\frac{x}{4}\right)^{\ga}
\Psi\left(\frac32,\ga+\frac32;(\ga+1)\frac{x}{4}\right) \ .
\label{rhogaPsi} 
\ee 
where $\Psi(a,b;z)$ is the Tricomi confluent
hypergeometric function (sometimes denoted $U(a,b;z)$). The
asymptotics for $x\to 0$ is easy to obtain as:
\begin{equation}
\label{Asympt x0}
  \rho_\ga(x)\sim
\frac{(\ga+1)^{1/2}\Gamma(\ga+1/2)}{\pi\Gamma(\ga+1)}
\frac{1}{\sqrt{x}}\ .
\end{equation}

A few comments are in order. First, the $\ga$-dependent prefactor
in (\ref{Asympt x0}) reproduces the correct WL asymptotic
behaviour from (\ref{MPc=1}) $\rho_{WL}(x)\sim
\frac{1}{\pi\sqrt{x}}$ for $x\to 0$ when $\ga\to\infty$. Secondly,
the WL inverse square root divergence at the origin is not
modified in the superstatistical generalisations, a feature that
is shared by the $\chi^2$-deformed WL \cite{AV}.

Next we turn to the large-$x$ asymptotics. Using only the leading
order term \cite{wolfram} 
\be 
\Psi(a,b;z)\ \sim \ z^{-a}\ \  \mbox{for  }|z|\to\infty\ , 
\ee 
we obtain immediately the following
result from eq. (\ref{rhogaPsi}) 
\be 
\rho_\ga(x) \sim
\frac{(\ga+1)^{\ga-1/2}}{4\sqrt{\pi}\Gamma(\ga+1)}
\exp\left[-(\ga+1)\frac{x}{4}\right] 
\left( \frac{x}{4}\right)^{\ga-\frac32}\ . 
\label{rhogaxoo} 
\ee
We illustrate the behaviour of the generalised semi-circle
density eq. (\ref{rhogaU}) in Fig. \ref{DensC1}. In order to
better resolve the behaviour at the origin, we map our spectral
density from the positive real axis to the full real axis by
defining
\be \vartheta_\ga(y)\ \equiv\ |y|\ \rho_\ga(y^2)\ ,
\label{Rmap}
\ee
that is to a normalised density on $\mathbb{R}$,
$\int_{-\infty}^\infty dy\vartheta_\ga(y)=1$. 
Furthermore, this map leads the {\it second} moment to be normalised 
$\int_{-\infty}^\infty dyy^2\vartheta_\ga(y)=1$.
In particular we
obtain for WL a semi-circle from eq. (\ref{MPc=1}),
\be
\vartheta_{W\!L}(y)=\frac{1}{2\pi}\,\sqrt{4-y^2}\ \  \mbox{on}\
[-2,2]\ .
\ee
In this squared-variables picture, a further, interesting feature of the
generalised spectral density eq. (\ref{rhoga}) appears when considering the
opposing limit $\ga\to 0$. There, the spectral density 
admits the following simplified form:
\be
\vartheta_{\ga\to
  0}(x)=|x|\frac{\Gamma\left(-\frac{1}{2},\frac{x^2}{4}\right)}{4\sqrt{\pi}} 
\ee
(where $\Gamma(x,y)$ is an incomplete Gamma function), which displays a cusp
at $x=0$. This feature often appears in spectral densities of complex networks
\cite{DM}. 

As a final remark in the standard WL (or WD) class the density also has an
exponential tail at finite-$N$, $\exp[-Nx^2]$. However, in the macroscopic
large-$N$ limit it disappears while in our deformed model such a  tail
persists.


\subsection{Generalised Mar\v{c}enko-Pastur law for $c<1$}\label{MP}

In this subsection we deal with the limit in which the matrix
\mbox{\bf X} remains rectangular, that is both $M$ and $N$ become
large with $N/M=c<1$ fixed. We follow the same steps as in
the previous subsection, first recalling the results for WL
that need to be incorporated into our model. It is known that in
the standard WL ensemble eq. (\ref{PWL}), the average density of
eigenvalues is given for large-$N$ as follows:
\begin{equation}
\label{MPc<1N}
\fl \lim_{N\gg 1}R_{W\!L,1}(\la)\ =\ \frac{\eta}{\pi\la}
\sqrt{\left(\la-\frac{N}{2\eta}X_-\right)\left(\frac{N}{2\eta}X_+-\la\right)}
\ ,\ \ \mbox{with}\ \ \la\in
\left[\frac{N}{2n}X_-,\frac{N}{2n}X_+\right]\ .
\end{equation}
Here we have defined the following bounds for later use \be X_\pm\
\equiv\ (c^{-\frac12}\pm1)^2\ ,\ \  \mbox{with}\ \ 0<c<1\ . \ee In
the limit $c\to1$ we recover from eq. (\ref{MPc<1N}) the
semi-circle eq. (\ref{densityWishart}) from the last section. An
$N$-independent density with mean one is again obtained by
rescaling with the mean eigenvalue position
$\langle\la\rangle_{W\!L}=\frac{M}{2\eta}$, and normalising
\bea \label{MPc<1} \fl
\rho_{MP}(x) &\equiv&  \lim_{N\to\infty}\frac1N\langle
\lambda\rangle_{W\!L} R_{W\!L,1}(x\langle \lambda\rangle_{W\!L})\
=\ \frac{1}{2\pi c x}
\sqrt{(x-cX_-)(cX_+-x)}\ ,\\
\fl&&\ \ \mbox{with}\ \ x\in[cX_-,cX_+]\ .\nn \eea This is called
Mar\v{c}enko-Pastur (MP) law \cite{MP}. Taking $c\to1$ we again recover
the $N$-independent semicircle eq. (\ref{MPc=1}).

Turning to our generalised model we now insert the large-$N$
result eq. (\ref{MPc<1N}) into eq. (\ref{RRrel}) and rescale with
the mean eq. (\ref{lagavev}) 
\bea \fl\rho_\ga(x) &\equiv&
\lim_{N,M\to\infty}\frac1N\langle \lambda\rangle_\ga
R_\ga(x\langle \lambda\rangle_\ga)
\label{rhoga2}\\
\fl&=&  \lim_{N,M\to\infty} \frac{\langle \lambda\rangle_\ga}{N}
\int_\mathcal{I}
d\xi\frac{\e^{-\frac{1}{\xi}}{Z}(\xi)\xi\ga}{\xi^{\ga+2-\frac\beta2
    NM}{Z}_\ga\pi x\langle \lambda\rangle_\ga}
\sqrt{\left(x\langle
\lambda\rangle_\ga-\frac{N}{2\xi\ga}X_-\right)\!\!
\left(\frac{N}{2\xi\ga}X_+-x\langle \lambda\rangle_\ga\right)}\nn
\eea
where
$\mathcal{I}\equiv\left[\frac{c}{x(\ga+1)}X_-,\frac{c}{x(\ga+1)}X_+\right]$
is the truncated integration range. Inserting the ratio eq.
(\ref{Zratio}) and substituting $\xi\to t=\xi x(\ga+1)/c$ we arrive at
the following 
\be \fl\rho_\ga(x)
=\frac{x^{\ga}}{2\pi\Gamma(\ga+1)}\left(\frac{\ga+1}{c}\right)^{\ga+1}
\int_{X_-}^{X_+}
\frac{dt}{t^{\ga+2}}\ \exp\left[-\frac{x(\ga+1)}{tc}\right]
\sqrt{(t-X_-)(X_+-t)}\ , \label{rhogac<1}
\ee
the second main result of this section. It is valid for any fixed
$\ga>0$ with $0<c<1$. As a check we can take the limit
$\ga\to\infty$, and a saddle point evaluation including the
fluctuations around the point $t_0=\frac{x}{c}$ leads back to the
MP density eq. (\ref{MPc<1}).

\begin{center}
\begin{figure*}[htb]
  \unitlength1.0cm
\centerline{  \epsfig{file=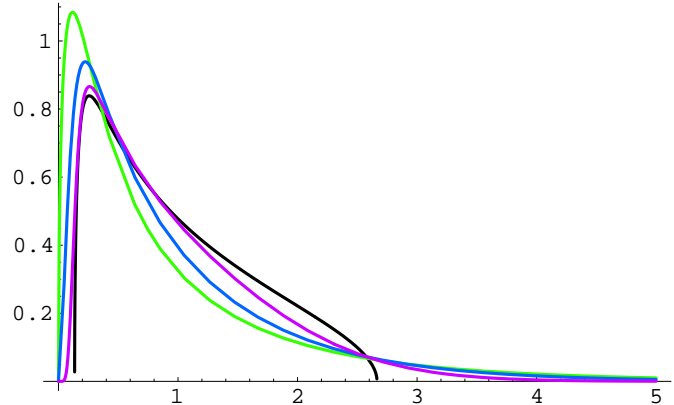,clip=,width=8.8cm}
}
  \caption{
    \label{DensCless1}
The generalised MP density eq. (\ref{rhogac<1}) for
$c=0.4$ and different values of $\ga=0.5,2,10$ (green, blue, violet
respectively). For comparison, the standard MP distribution eq.
(\ref{MPc<1}) is plotted as well (black). }
\end{figure*}
\end{center}

\begin{center}
\begin{figure*}[htb]
  \unitlength1.0cm
  \epsfig{file=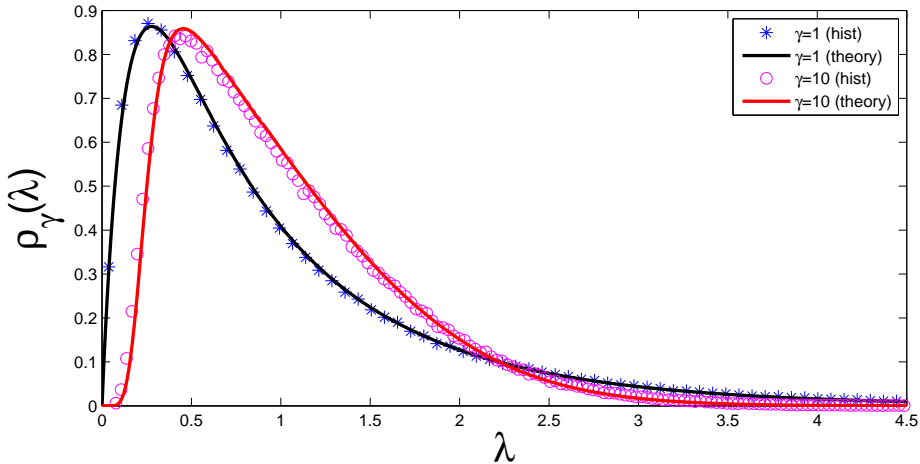,clip=,width=13.8cm}
  \caption{
    \label{DensCless1num}
Numerical check for $\beta=1$:
Histogram of eigenvalues for $N=10$, $M=40$ and $\gamma=1,10$ (blue and
magenta, respectively), together with 
the theoretical result eq. (\ref{rhogac<1}) for $N=\infty$ 
(black and red solid lines). The average is obtained over $R=50000$ samples.}
\end{figure*}
\end{center}

Next we turn to the asymptotic analysis for $x\to 0$ and
$x\to\infty$. For small values of $x$ we simply obtain \be
\rho_{\ga}(x)\sim D_{\ga} x^{\ga} \ee
where the constant $D_\ga$ easily follows from eq. (\ref{rhogac<1}).
For $x\to\infty$, we obtain:
\be
\rho_{\ga}(x)\sim
\frac{\sqrt{X_{+}-X_{-}}}{4\sqrt{\pi}
\Gamma(\ga+1)(X_{+})^{\ga-1}}\left(\frac{\ga+1}{c}\right)^{\ga-1/2}
x^{\ga-3/2}\rm{e}^{-\frac{(\ga+1)}{c X_{+}}x}\ .
\ee
As a check, we can recover the correct asymptotical behaviour
(\ref{rhogaxoo}) when $c\to 1$ (in this case $X_{-}\to 0$ and
$X_{+}\to 4$).
Note that the decay for large arguments is the same as for the
generalised semi-circle eq. (\ref{rhogaxoo}), in terms of variables $x/(cX_+)$.
An example for eq. (\ref{rhogac<1}) is shown in Fig. \ref{DensCless1}.
Such exponentially decaying correlations occur in exponentially growing
complex networks, apart from more frequent power law decays \cite{DM}.
In addition to our analytical checks comparing to WL and $c=1$ we have also
performed numerical simulations, generating matrices 
in our generalised class with the algorithm
given in \ref{numerical}. 
In Fig. \ref{DensCless1num} we compare the numerical results for the spectral density at finite $N,M$ 
with the theoretical prediction (valid at infinite $N$) eq. (\ref{rhogac<1}).
We find an excellent agreement already for moderate
$N$ and $M$ ($N=10$, $M=40$).

\section{Level Spacing from a Wigner Surmise}
\label{spacing}

The exact computation of the level spacing distribution for finite matrix size $N$
(and in the microscopic large-$N$ limit in the bulk) 
is mathematically quite nontrivial, even
in the simplest case of the WD ensembles, see e.g. \cite{mehta}. 

Therefore long ago Wigner came up with the idea to compute the exact spacing
distribution in the WD class for the simple $2\times 2$ case, obtaining 
\be
\mathrm{P}_{W\!D}^{(\beta)}(s)= a_\beta s^\beta \exp[-b_\beta s^2]\ .
\label{WD}
\ee
The $\beta$-dependent constants $a_\beta, b_\beta$ easily follow by fixing the 
norm and first moment to unity (see e.g. in \cite{guhr}). 
Below we will mostly need $\beta=1$ with 
$a_1=\pi/2$, and $b_1=\pi/4$.
The result eq. (\ref{WD}) turns out to be an excellent approximation of
the true result for moderately large $N$, see e.g. Fig. 1.5 in \cite{mehta}, and we aim at a similar
approximate representation for the spacing distributions in superstatistical ensembles. 

However, before undertaking this more complicated task,
we ask the simpler question: does a Wigner surmise using $N=2$ work well in the
unperturbed WL ensemble? 
We have not seen this issue discussed in depth in the literature (see however \cite{plerou} section V and references therein)
and we feel it is appropriate to spend a few words on it in the next section.

\subsection{A Wigner surmise in WL?}

Starting from eq. (\ref{PWLev}) 
for $N=2$ with weight $\exp[-n\beta\la]$ ($n=1/2$ for standard normal entries),
we can compute the exact spacing
distribution ${\cal  P}_{\bar{\nu}}(s)$ in WL by integrating over both eigenvalues with the constraint
$\delta(\la_2-\la_1-s)$. Since this was already done in \cite{AV} we can just quote the result:
\begin{equation}\label{GapDistribution}
{\cal  P}^{(\beta)}_{\bar{\nu}}(s)=C
  s^{\beta+\bar{\nu}+1/2}K_{1/2+\bar{\nu}}(n\beta s)\ ,
\end{equation}
where $K_\mu(x)$ is a modified Bessel function,
$\bar{\nu}\equiv\frac{\beta}{2}(M-N+1)-1=\frac{\beta}{2}(M-1)-1$, and the constant $C$
is given by:
\begin{equation}\label{CWL}
  C=\left(2^{-1/2+\beta+\bar{\nu}}(n\beta)^{-3/2-\beta-\bar{\nu}}
\Gamma\left(\frac{1+\beta}{2}\right)
\Gamma\left(1+\bar{\nu}+\frac{\beta}{2}\right)\right)^{-1}
\end{equation}
to ensure a normalisation to unity. 
The first moment can be normalised by
computing 
\be
d\equiv \int_0^\infty ds\,s {\cal  P}^{(\beta)}_{\bar{\nu}}(s)\ ,
\label{mean}
\ee
and then defining 
\be
\hat{{\cal  P}}^{(\beta)}_{\bar{\nu}}(s)
\equiv d\ {\cal  P}^{(\beta)}_{\bar{\nu}}(sd)\ .
\label{1stmomd}
\ee

The result eq. (\ref{1stmomd}) is obviously different from
eq. (\ref{WD}), even for $M=N=2$. Can eq. (\ref{1stmomd}) be a better approximation than eq. (\ref{WD})
for the true WL spacing at $N$ finite but large?

In order to compare, let us pick $N=M$ and $\beta=1$ or $2$ 
implying $\bar{\nu}=-\frac12$ or 0 respectively. 
The resulting Bessel-$K$ function simplifies only for 
half-integer
index, $K_{1/2}(x)=\sqrt{\pi/2x}\,e^{-x}$, and we obtain
\bea
{\cal  P}^{(\beta=1)}_{\bar{\nu}=-\frac12}(s)&\sim& s K_0(s)
\label{Pbeta=1}\ ,\\
{\cal  P}^{(\beta=2)}_{\bar{\nu}=0}(s)&\sim& s^2 \exp[-2s]\ .
\label{Pbeta=2nu=0}
\eea
After normalising the first moment, neither case matches with eq. (\ref{WD})  
and for $N\neq M$ (or $\beta=4$) the difference is
even more pronounced. 

\begin{center}
\begin{figure*}[htb]
  \unitlength1.0cm
  \epsfig{file=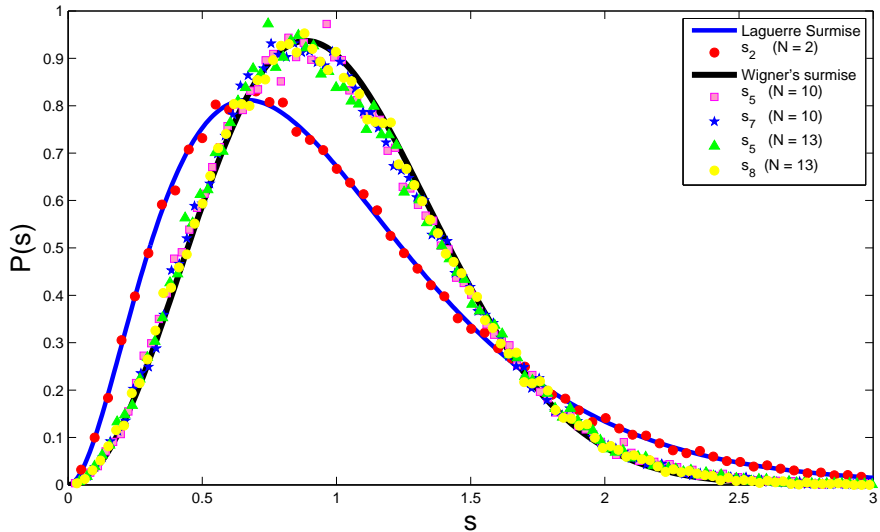,clip=,width=13.8cm}
  \caption{
    \label{surmisecheck}
Numerical check of the WL and WD surmise for $\beta=1$ and $N=M$: 
$N=2$ vs. $N=10$ $(13)$, for various locations in the bulk: $k=5,7,8$.
}
\end{figure*}
\end{center}

In order to check which spacing distribution is a better approximation for larger $N$ instances, we performed
numerical simulations using the Dumitriu-Edelman tridiagonal algorithm \cite{dumitriu} 
for the case $\beta=1$ and $N=M$, with the result
shown in Fig. \ref{surmisecheck}. Before commenting on this we mention in
passing our procedure. Usually, the numerical quantity to be compared to the surmise
is the so-called \emph{nearest-neighbour spacing distribution}, i.e. the normalized histogram of \emph{all}
the spacings among consecutive eigenvalues in the bulk after unfolding.
This procedure is often time-consuming and special care is needed to avoid spurious effects in the analysis (see
\cite{plerou,unfolding} and references therein for a detailed discussion on unfolding procedures).

We can thus resort to the following,
equivalent method to extract the \emph{individual} spacing at a given location $k$ in the spectrum (see \cite{Mspace}) as:
\be
s_{k}\equiv \frac{\la_{k}-\la_{k-1}}{\langle\la_{k}-\la_{k-1} \rangle}\ .
\ee
The average $\langle\cdot\rangle$ is taken over many samples 
and obviously $\langle s_k\rangle=1$. For a given $N$ we
have picked different values of $k$ to check that our results do not depend on
the position in the bulk.

While the numerical histogram for $N=2$ follows the exact result for the spacing
from eqs. (\ref{Pbeta=1}), for increasing $N$ it quickly converges towards
the WD surmise eq. (\ref{WD}). We have checked that 
this holds when choosing $N\neq M$ as well.

What is the explanation?
The reason lies in the well known fact that in the bulk both 
in the WD and WL
  classes the correlations are governed by the Sine kernel (for a rigorous proof, see \cite{universal}). 
The exact spacing distribution can then be expressed in terms of Fredholm
  eigenvalues of this kernel, see. e.g. \cite{mehta}, which happens to be well
  approximated by the WD surmise. This is why the WD surmise applies to both
  ensembles. In contrast the 
$N=2$ WL surmise is not a good approximation,
and so the surmise does not work
  here at all. In the next subsection, we seek for a generalised Wigner's surmise holding in the bulk of our superstatistical model.

\subsection{Level spacing for our generalised model} 

How should an approximate spacing distribution for the
generalised WL model with an inverse $\chi^2$-distribution be constructed? Because of the failure of an $N=2$ surmise in WL
 (as we just explained), 
we should \emph{not} expect that inserting it into our integral transform will work in the superstatistical case. 
Instead, we will follow a more heuristic approach confirmed by various 
numerical checks. 

Rather that following the procedure described in subsect. \ref{univ} 
we directly
start from an integral transform of eq. (\ref{WD}) but with a $\xi$-dependent
variance\footnote{The quadratic rather than linear power in $\xi$ in the
  exponent can
  be motivated by a change of variables in eq. (\ref{PWLev})  $\rm{e}^{-\xi\la}\to
  \rm{e}^{-\xi^2\la^2}$ from the WL to the Gaussian WD weight.} 
\be
\mathrm{P}_{W\!D}^{(\beta)}(s;\xi)= 2 \xi^{\beta+1}
{\Big(\frac\beta2\Big)^{\frac{\beta+1}{2}}}
{\Gamma\Big(\frac{\beta+1}{2}\Big)^{-1}} s^\beta 
\exp[-\beta\, \xi^2s^2/2]\ .
\label{WDxi}
\ee
Eq. (\ref{WDxi})
is normalised to one (the first moment will be normalised below).
Because of $\int d\xi f(\xi)=1$ the following folded surmise for our 
generalised WL is also normalised:
\be
\fl P_{\ga}^{(\beta)}(s)\equiv 
\int_0^\infty d\xi
\xi^{-\ga-2} e^{-\frac1\xi} \mathrm{P}_{W\!D}^{(\beta)}(s;\xi)=
C_\ga s^\beta \int_0^\infty d\xi
\xi^{-\ga-1+\beta} \exp\Big[-\frac1\xi-\frac12\beta\,\xi^2s^2\Big]\ ,
\label{newsurmise}
\ee
with 
\be
C_\ga =2{\Big(\frac\beta2\Big)^{\frac{\beta+1}{2}}}{\Gamma\Big(\ga+1\Big)^{-1}
\Gamma\Big(\frac{\beta+1}{2}\Big)^{-1}}.
\ee

\begin{center}
\begin{figure*}[h]
  \unitlength1.0cm
  \epsfig{file=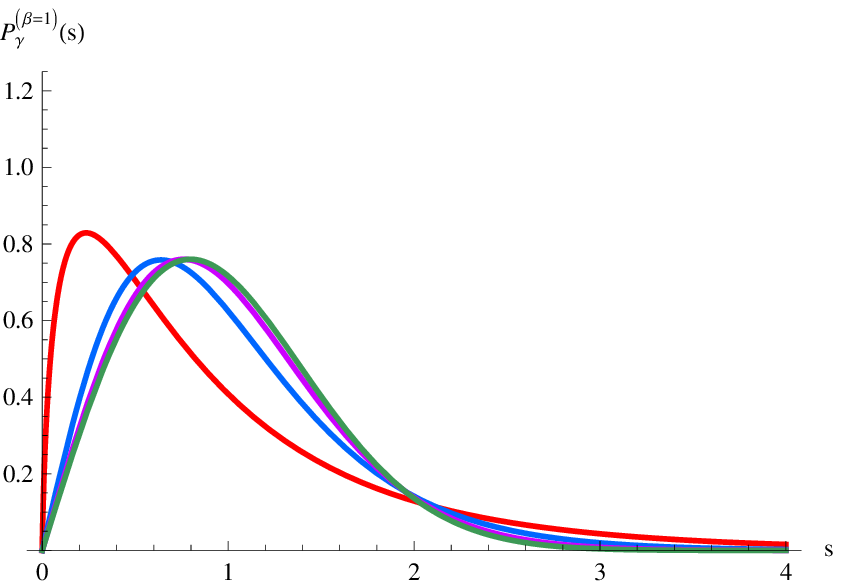,clip=,width=5.1cm}
  \epsfig{file=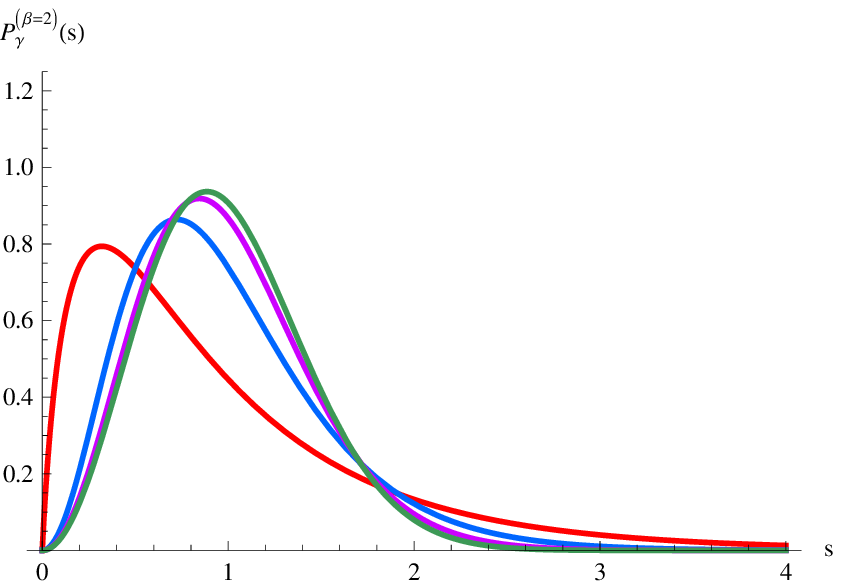,clip=,width=5.1cm}
  \epsfig{file=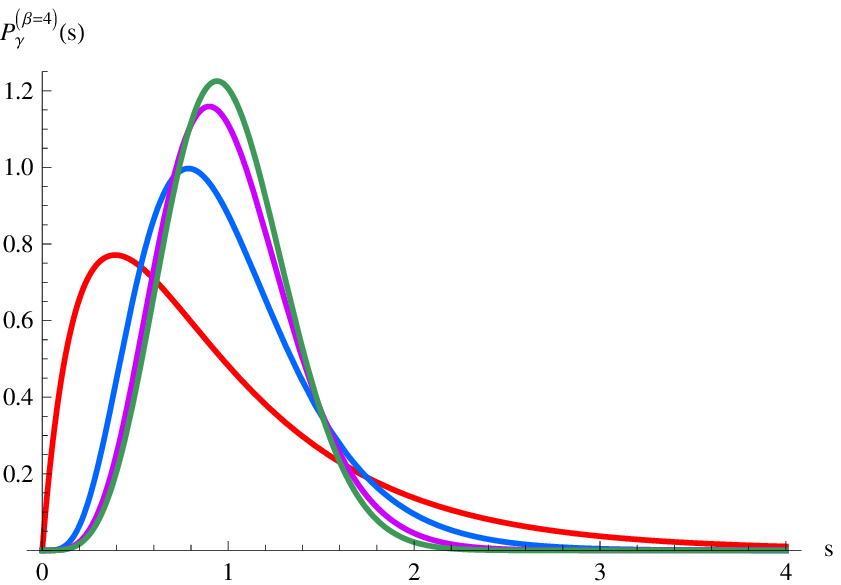,clip=,width=5.1cm}
  \caption{
    \label{spacingallb}
Level spacing distribution 
eq. (\ref{gap2}) at $\beta=1,2,4$ (from left to right) and
different values of $\ga=1,10,50$ (red, blue, violet
respectively). We have also included the WD surmise eq. (\ref{WD}) in green for comparison.
}
\end{figure*}
\end{center}
To normalise the first moment we compute 
\be
d_\ga\equiv \int_0^\infty ds\,s P_{\ga}^{(\beta)}(s)=
\frac{(\ga+1)\Gamma\Big(\frac{\beta}{2}+1\Big)}
{\Big(\frac{\beta}{2}\Big)^{\frac12}\Gamma\Big(\frac{\beta+1}{2}\Big)}
\ee
and obtain the approximate spacing distribution in our generalised WL with
an inverse $\chi^2$-distribution:
\be
\hat{P}_{\ga}^{(\beta)}(s)\equiv d_\ga  P_{\ga}^{(\beta)}(sd_\ga)\ .
\label{gap2}
\ee
Eq. (\ref{gap2}) is the main result of this subsection.
Given the known universality of the (approximate) eq. (\ref{WD}), our new
spacing distribution will also be universal under a large class of
deformations as discussed in subsection \ref{univ}. The integral can be evaluated
in terms of hypergeometric functions, but we prefer to keep the integral
form for simplicity. 
We have checked explicitly that in the limit $\ga\to\infty$ we correctly
reproduce eq. (\ref{WD}) valid for WL $(N\gg 2)$.

Eq. (\ref{gap2}) is plotted for all three values of $\beta=1,2,4$ and
various $\ga$ in Fig. \ref{spacingallb}, including the limiting WD
distribution eq. (\ref{WD}).

Our new surmise eq. (\ref{gap2}) has the following asymptotic
behaviour. The level repulsion at short distances ${s\to0}$ 
is given by
\be
\hat{P}_{\ga}^{(\beta)}(s)\ \sim\  \kappa_1 s^{\min(\beta,\ga)}\ ,
\ee

At large distances ${s\to\infty}$ 
however we get a new behaviour in terms of a stretched
exponential 
\be
\hat{P}_{\ga}^{(\beta)}(s)\ \sim\ \kappa_2 s^{\frac13(\beta+2\ga-1)}\exp[-\alpha_\ga s^{\frac23}]\ ,
\ee
which can be obtained after changing variables and making a saddle point
approximation ($\alpha_\ga=\frac32(\beta d_\ga^2)^{1/3}$). In both formulae, we have omitted 
the exact form of the $\ga$-dependent prefactors $\kappa_{1,2}$.

\begin{center}
\begin{figure*}[htb]
  \unitlength1.0cm
\centerline{
  \epsfig{file=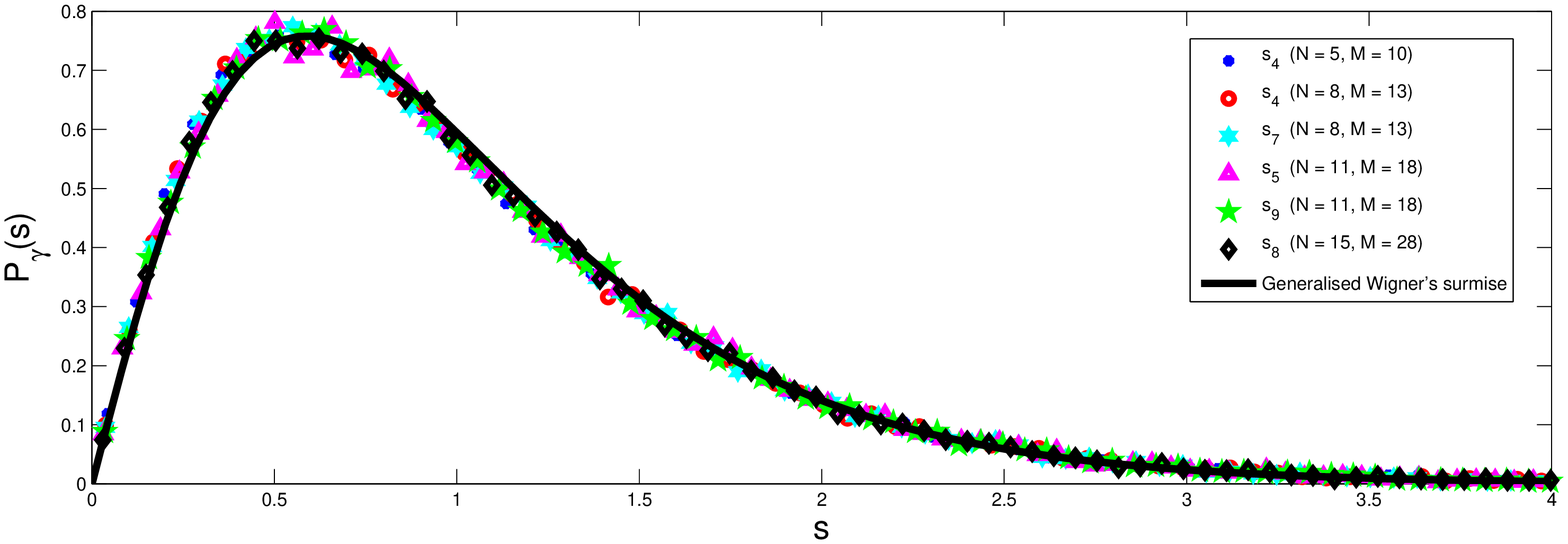,clip=,width=13.8cm}} 
\centerline{
  \epsfig{file=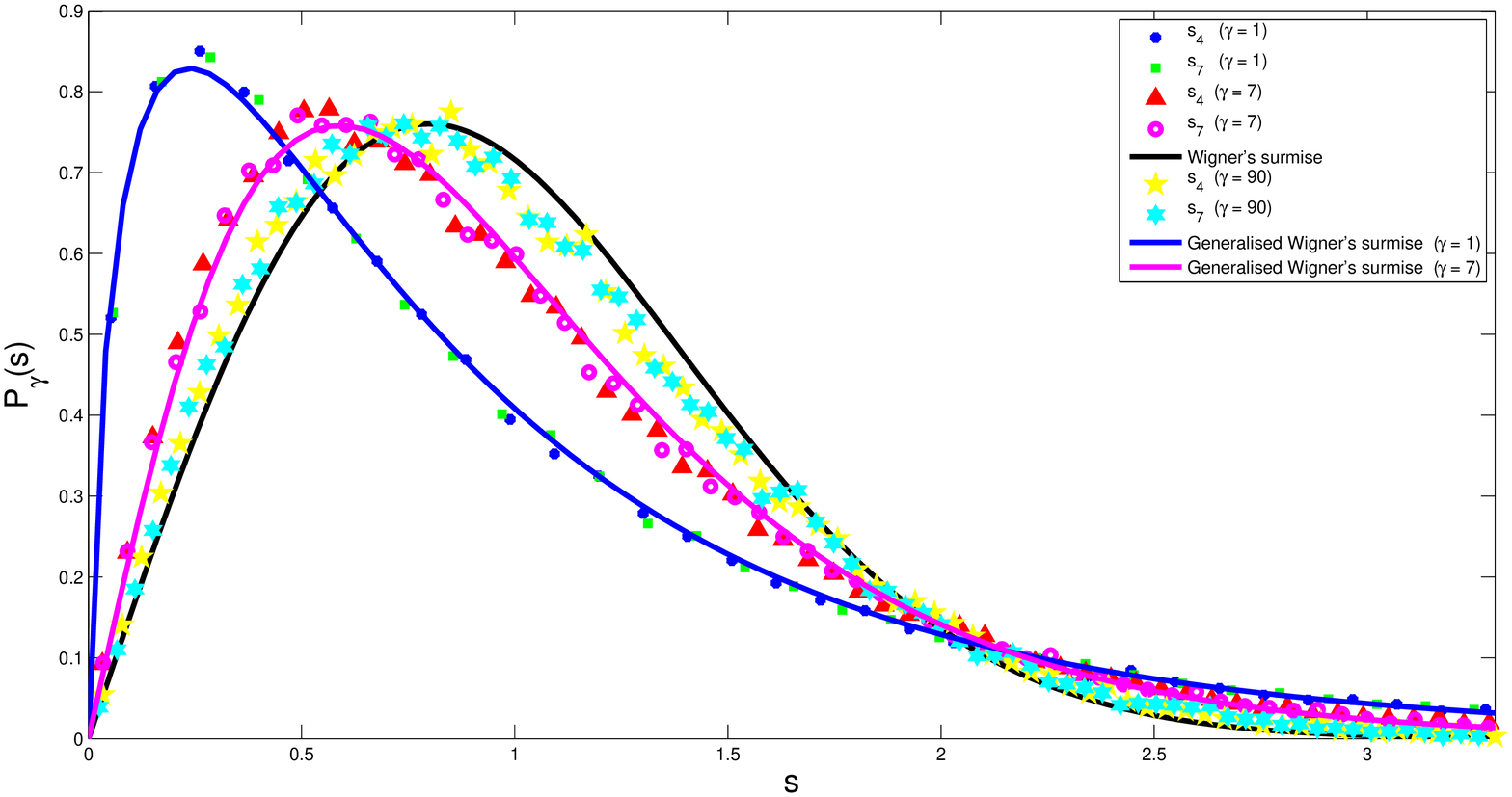,clip=,width=13.8cm}} 
  \caption{
    \label{spacingdata}
Comparison to numerical simulations; top: for fixed $\ga=7$ a comparison
between eq. (\ref{gap2}) 
and numerically generated spacings for various combinations of $N,M$ and locations $k$
within the bulk; bottom: for fixed $N=10,\ M=15$   eqs. (\ref{gap2}), (\ref{WD})
and numerically generated spacings for various values of $\ga$ and locations $k$. 
}
\end{figure*}
\end{center}

To test our new surmise, we have performed numerical simulations of the individual spacing distribution in the bulk 
for the most
relevant case $\beta=1$ and various values of $\gamma,N$, $M$ and $k$ (the location within the spectrum). 
Our findings are summarised in
Fig. \ref{spacingdata}, displaying an excellent agreement between the numerical histogram and our
surmise. 

In Fig. \ref{spacingdata} (top), we keep $\ga$ fixed to the value $\ga=7$ 
and vary $N,M$ and the location $k$ within the spectrum. The histogram of the individual $k$\textit{th} spacing
reveals the independence of the
spacing on the location $k$  and on $N$ or $M$ (sufficiently greater than $2$). 
In Fig. \ref{spacingdata} (bottom), we keep $N,M$ fixed to the values $(10,15)$ respectively.
Increasing $\ga$ ($\ga=1,7,90$) we can nicely see the convergence of both the
numerical histogram and theoretical curves towards the WD surmise eq. (\ref{WD}). 


\section{Applications and Discussion}\label{apply}

Before comparing to actual data, let us put our results into the context of
other models and highlight their features. 

From a RMT point of view, the first classification issue is whether or not
the considered ensemble is invariant. Preserving invariance leads to a
complete solvability for all spectral correlators in our approach as well as
in \cite{AV}. On the other hand, non-invariant models as in
\cite{burda,biroli,pato} permit generically to compute only the macroscopic
density, either implicitly or explicitly. 
A second issue is that of RMT universality classes, 
and we argued in section
\ref{univ} that our macroscopic density is universal in a
weak sense under (non-invariant) deformations using independent random
variables. 

A third and important issue is to classify 
how the spectral density decays for large
arguments, even though this may be difficult to appreciate from a small set of
data. We have exemplified spectral densities  with a power law and 
exponential decay using a $\chi^2$- and inverse $\chi^2$-distribution
respectively. 

About the first issue above, we mention that the non-invariant ensemble \cite{biroli} using a multivariate
student distribution - a product of Gaussian and inverse $\chi^2$-distributed
random variables - leads to a density with power law decay. 
Another example is the deformed Gaussian Orthogonal
Ensemble (GOE) \cite{pato} having an exponential 
tail as in our more general eq. (\ref{rhogaxoo}),
for the special case of $\ga=2$ (and $c=1$). This model is non-invariant as
well and can be derived from another entropy maximisation procedure \cite{CJH}. 
\begin{center}
\begin{figure*}[h]
  \unitlength1.0cm
\centerline{  \epsfig{file=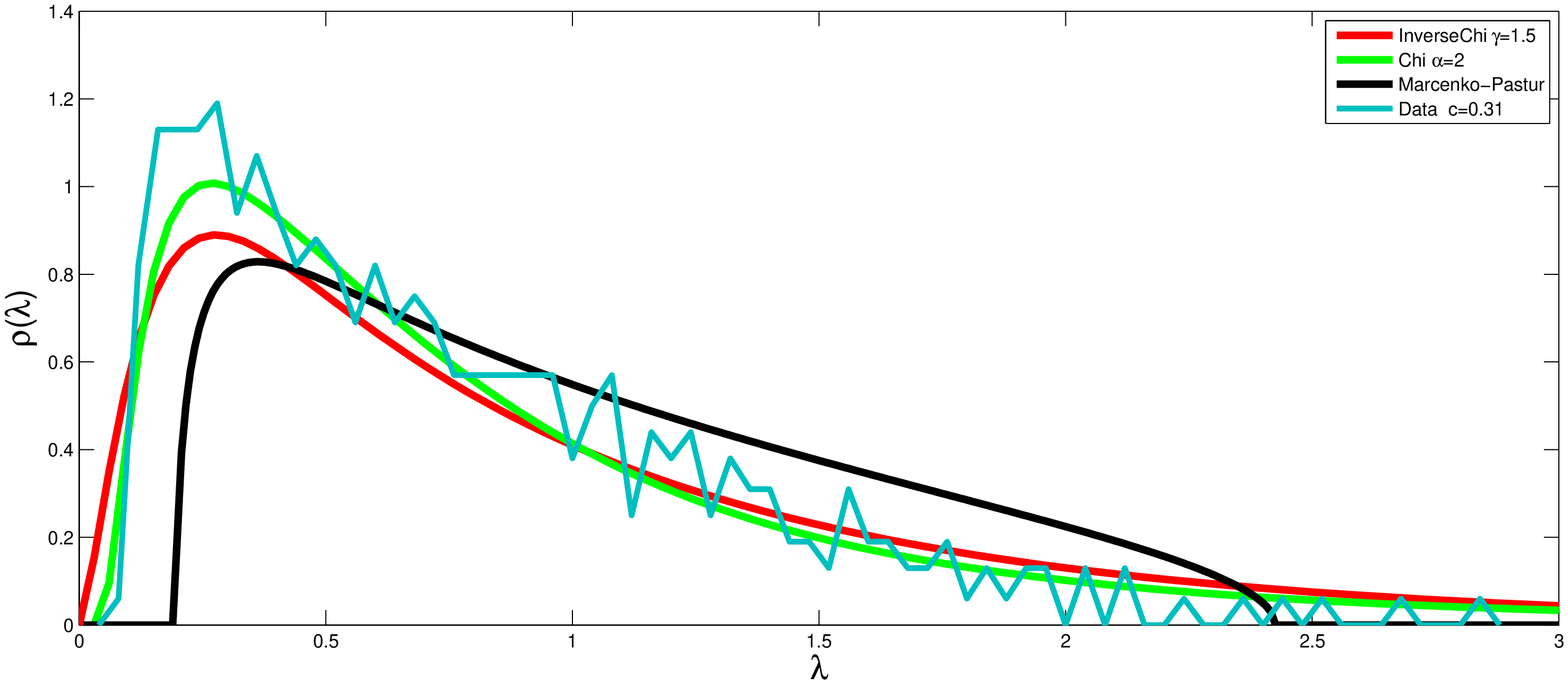,clip=,width=13.8cm}}
\centerline{  \epsfig{file=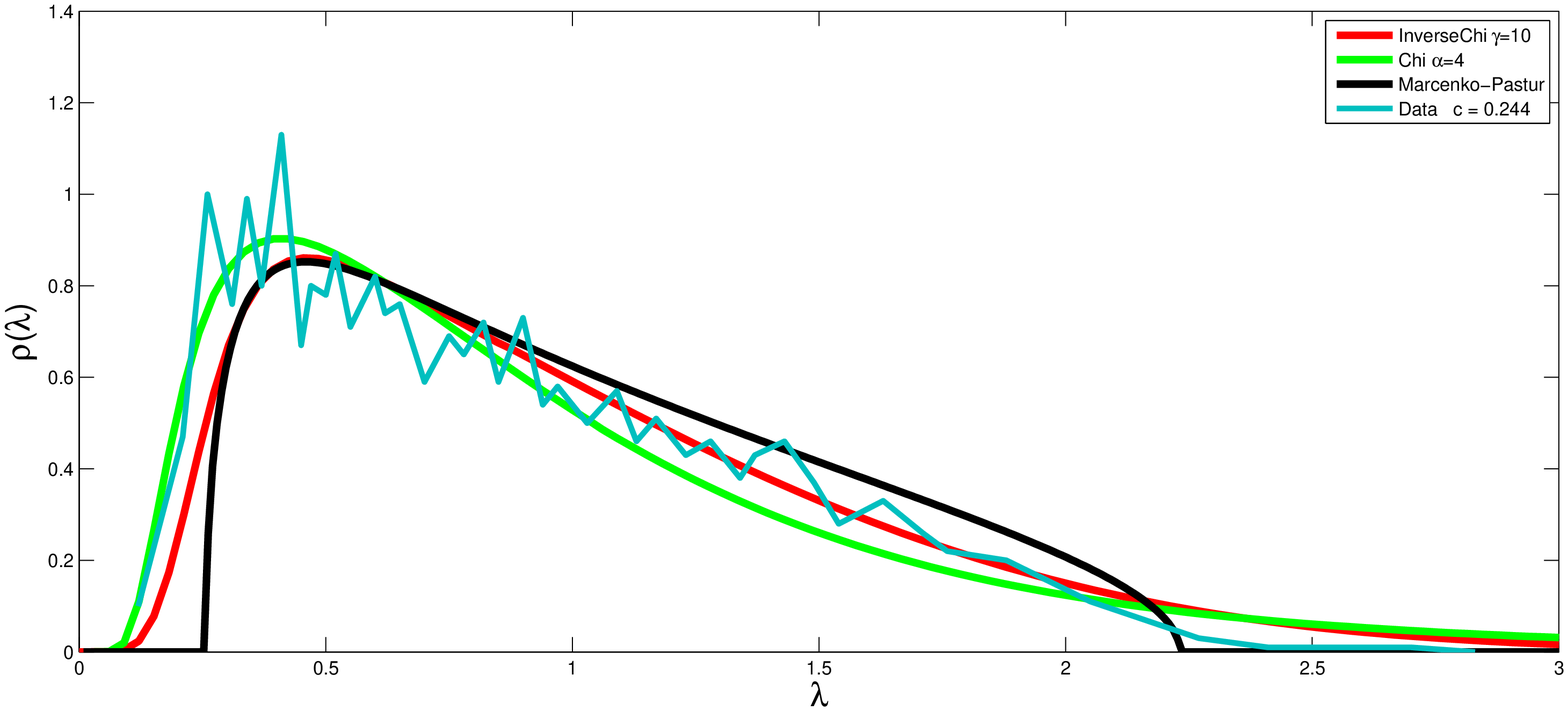,clip=,width=13.8cm}}
  \caption{
    \label{data}
A comparison to 2 sets of spectral densities from financial covariance
matrices: top S\&P500 price fluctuations of $N=406$ assets of $M=1309$ days
between 1991-1996 \cite{laloux} (see also \cite{ZJ} Fig 1.), and 
bottom daily price exchanges 
of the Johannesburg Stock Exchange from Jan 1993-Dec 2002 \cite{wilcox}.
In both cases a few higher eigenvalues are omitted.
The smooth curves are the spectral densities for the given $c<1$ from a
$\chi^2$-distribution (green) \cite{AV}, inverse $\chi^2$-distribution (red)
eq. (\ref{rhogac<1}) and for comparison, the standard MP distribution eq.
(\ref{MPc<1}) (black). 
}
\end{figure*}
\end{center}
The most natural setting to apply WL or its generalisations is in time
series analysis. In Fig. \ref{data} we compare to eigenvalues from financial covariance
matrices for 2 different sets of data \cite{ZJ,wilcox}\footnote{We kindly thank
the authors for permission to use the data from their papers.}, exploiting superstatistical models with
$\chi^2$- and inverse $\chi^2$-distribution (displaying power law and 
exponential decay respectively). While in the top plot the former gives a better
fit, in the lower plot our new eq. (\ref{rhogac<1}) for the inverse
$\chi^2$-distribution gives at least a comparable
if not better fit to the data. We also compare to the standard 
MP distribution eq. (\ref{MPc<1}), which appears to give clearly a less good fit. 
Because in financial data there is no easy underlying physical principle for
identifying extensive variables and thus 
the distribution class to be applied, our comparison
must be heuristic, deciding case by case which distribution gives the best fit.

A similar approach has been taken in a comparison with partly
chaotic billiards \cite{sust2}. Here the microscopic spacing
distribution from various generalised WD classes 
is compared to the data, and the inverse $\chi^2$-distribution 
gives the best fit.
Other examples where the inverse $\chi^2$-distribution has been recently
identified is in turbulent flows \cite{VdSB}, although not in a RMT setting.

A very recent example of an application of a superstatistical RMT 
to complex networks was given in
\cite{CJH}. Here the deformed GOE model \cite{pato} within the class of 
exponentially
decaying densities is compared to local statistics of the adjacency matrix. Due
to its non-invariance the microscopic RMT predictions there are obtained from
simulations. It would be very interesting to calculate these correlation
functions analytically in the framework of our model with a more general
exponential decay, but this is going beyond the scope of this article.

\section{Conclusions}\label{conclusio}

Using the recently proposed methods of superstatistics, we 
provided a systematic framework for generalising the
Wishart-Laguerre ensembles of random covariance matrices, 
retaining the exact solvability of the original model. This is
achieved by allowing the ensemble parameter - the inverse variance
of the data matrix elements - to fluctuate from one sample to
another according to a certain distribution $f$, and then
averaging over $f$. We have given a 
new compact expression for the
distribution of matrix elements in the particular case where the
ensemble parameter has an inverse $\chi^{2}$-distribution. This distribution has
appeared in the literature when modelling the volatility of financial markets and thus it is of 
interest for the analysis of large arrays of data. 
Averaging over this distribution, we are able to express the
spectral statistics of the generalised WL ensembles as an integral
over the corresponding statistics for the standard WL ensemble.
This is the key ingredient to solve the model exactly, 
for finite-$N$ as well as in both the
macroscopic and microscopic limits. 
In our model we have full control over the interplay between the deformation
parameter $\ga$, the matrix size $N$ and their respective asymptotic limits.

We have computed exactly
several spectral quantities, first deriving a generalised
semi-circle and generalised Mar\v{c}enko-Pastur density in the
macroscopic large-$N$ limit for square and rectangular data
matrices respectively: in both cases, we obtain an exponential
tail. Secondly, after discussing in detail the level spacing distribution in WL ensembles
and the applicability of the standard Wigner's surmise, we determined the microscopic level spacing
distribution for all three $\beta$ using a new, $\ga$-dependent surmise, which exhibits a good agreement
with numerical simulations.

Our findings are illustrated via an application to
financial covariance matrices where we make a comparison between fits resulting from 
a $\chi^2$- and an inverse $\chi^2$-distribution. 
It would be very interesting to find further applications, in particular to
time series of other kinds of data. 
\\

\noindent \underline{Acknowledgements}:
This work started while one of us (A.Y.A-M) was in a short visit to Brunel
University. The hospitality of the members of Department of Mathematical
Sciences is acknowledged. A.P. Masucci is thanked for providing a reference.
Financial support by EPSRC
grant EP/D031613/1, European Network ENRAGE MRTN-CT-2004-005616
(G.A.), and European Union Marie Curie Programme NET-ACE (P.V.) is
gratefully acknowledged. We are very grateful to R. K\"uhn for discussions
and  
kindly sharing his data.\\

\begin{appendix}

\section{Numerical Simulations}\label{numerical}

Since it is not quite trivial to generate matrices with a non-standard
Gaussian distribution we briefly describe the algorithm in detail.
We focused on matrices with real elements ($\beta=1$) 
as an example, being the most relevant case for applications.

The algorithm used is the following:
\begin{enumerate}
 \item Draw a random variable $\xi$ from the inverse-$\chi^2$ distribution

 $f(\xi)=\xi^{-\ga-2}e^{-1/\xi}/\Gamma(\ga+1)$; 
 \item Generate a random $M\times N$ matrix $\mathbf{X}$ whose entries are
 normal variables with vanishing mean, and variance $\sigma^2=1/(2\ga\xi)$; 
 \item Generate the covariance matrix
 $\mathbf{\tilde{W}}=\mathbf{X}^T\mathbf{X}$; 
 \item Diagonalise $\mathbf{\tilde{W}}$, obtaining its $N$ positive
 eigenvalues $\{\mu_1^{(\ell)},\ldots,\mu_N^{(\ell)}\}$ ($\ell$ stands for the $\ell$\textit{th} sample); 
 \item Store:

- in the matrix row $\mathcal{V}_{\ell,j}$ the rescaled eigenvalues
 $\{\lambda_j^{(\ell)}\}$ ($j=1,\ldots,N$) (where $\lambda_j^{(\ell)}=N\mu_j^{(\ell)}/\sum_{k=1}^N
 \mu_k^{(\ell)}$) 

- in the matrix row $\tilde{\mathcal{S}}_{\ell,j}$ the bare spacings
 $\tilde{s}_j^{(\ell)}=\mu_{j}^{(\ell)}-\mu_{j-1}^{(\ell)}$ ($j=2,\ldots,N$); 
 \item Iterate the procedure $R$ times, so that $\ell=1,\ldots,R$;
 \item Plot i) a normalised histogram of all the entries $\mathcal{V}_{\ell,j}$ (average spectral density)
and ii) given a certain $k$ between $2$ and $N$, a normalised histogram 
of all the entries of $\tilde{\mathcal{S}}$ in column $k$, normalised by the mean $(1/R)\sum_{r=1}^R \tilde{\mathcal{S}}_{r,k}$
(individual spacing distribution at location $k$).
\end{enumerate}
The plots of the normalised histograms are then compared with the theoretical
results for finite or infinite $N$ respectively (see Figs. \ref{DensCless1num},\ref{surmisecheck} and \ref{spacingdata}).


\end{appendix}

\end{document}